\documentstyle[12pt,epsfig]{article}
\newlength{\dinwidth}
\newlength{\dinmargin}
\newcommand{\ba}{\begin{array}}
\newcommand{\ea}{\end{array}}
\newcommand{\ben}{\begin{equation}}
\newcommand{\een}{\end{equation}}
\newcommand{\bea}{\begin{eqnarray}}
\newcommand{\eea}{\end{eqnarray}}
\newcommand{\gsim}{\mathrel{\mathop{\kern 0pt \rlap
  {\raise.2ex\hbox{$>$}}} \lower.9ex\hbox{\kern-.190em $\sim$}}}

\def\nn{\nonumber}
\def\half{{1\over 2}}
\def\e2{{\epsilon \over 2}}
\def\p{\prime}

\def\ui{u^{in}}
\def\uis{u^{in *}}
\def\uo{u^{out}}
\def\uos{u^{out *}}
\def\a{\alpha}

\def\b{\beta}

\def\ai{a^{in}}
\def\ao{a^{out}}
\def\aid{a^{in \dagger}}
\def\aod{a^{out \dagger}}
\def\wb{\bar \omega}
\def\w{\omega}
\def\lt{\tilde \lambda}

\def\a{\alpha}
\def\b{\beta}
\def\wb{\bar w}

\def\ui{u^{in}}
\def\uo{u^{out}}

\begin{document}
\thispagestyle{empty}
\addtocounter{page}{-1}
\begin{flushright}
UK-04-08\\
MCTP-04-21\\ {\tt hep-th/0403275}
\end{flushright}
\vspace*{1.3cm}
\centerline{\large \bf Particle Production in Matrix Cosmology}
\vspace*{1.2cm} 
\centerline{\bf Sumit R. Das${}^{a}$, Joshua L. Davis${}^{b}$, Finn Larsen${}^{b}$, and
Partha Mukhopadhyay${}^{a}$}
\vspace*{0.8cm}
\centerline{\it ${}^a$Department of Physics and Astronomy}
\vspace*{0.2cm}
\centerline{\it University of Kentucky, Lexington, KY-40506, U.S.A.}
\vspace*{0.6cm}
\centerline{\it ${}^b$Michigan Center for Theoretical Physics}
\vspace*{0.2cm}
\centerline{\it University of Michigan, Ann Arbor, MI-48109, U.S.A.}
\vspace*{1cm}
\centerline{\tt das@pa.uky.edu, joshuald@umich.edu, larsenf@umich.edu, partha@pa.uky.edu}
\centerline{\tt }
\vspace*{1.5cm}
\centerline{\bf Abstract}
\vspace*{0.5cm}
We consider cosmological particle production in 1+1 dimensional string
theory. The process is described most efficiently in terms of anomalies, 
but we also discuss the explicit mode expansions. In matrix cosmology
the usual  vacuum ambiguity of quantum fields in time-dependent backgrounds
is resolved by the underlying matrix model.
This leads to a finite energy density for the "in" state which cancels the
effect of anomalous particle production.
\vspace*{1.1cm}

\baselineskip=18pt
\newpage

\section{Introduction}
The spontaneous production of quantum particles in curved backgrounds is a 
profound phenomenon which has inspired progress in fundamental physics for 
several decades. In particular, cosmological particle production is, according to 
inflationary cosmology, the origin of all observable structure in the universe. It
is clearly important to understand particle production better in a context where 
the quantum nature of gravity is taken fully into account. The purpose of this 
paper is to attempt this in the case of the matrix model for two 
dimensional string theory. 

In the usual treatment of quantum fields in curved space-time cosmological
particle production is the tangible consequence of an ambiguous vacuum: 
the vacuum state defined with respect to modes which are natural at early times 
typically contains particles when analyzed with respect to modes which are 
natural at late times (for a review see \cite{Birrell:ix}). 
This is quite similar to Hawking radiation from black holes \cite{Hawking:sw}. 
In string theory, holographic representations of gravity in terms of gauge theories 
have led to significant progress in our understanding of Hawking radiation
\cite{Strominger:1996sh, Callan:1996dv, Horowitz:1996fn, 
Das:1996wn, Maldacena:1997re,
Witten:1998qj, Gubser:1998bc, Witten:1998zw}.
In the gauge theory description, there is a preferred time and so a preferred 
vacuum; and Hawking radiation results arises through the 
normal quantum mechanical decay of a 
highly degenerate initial state 
\cite{Callan:1996dv,Das:1996wn,Dhar:1996vu, Maldacena:1996ix}. 
It is natural to 
expect that string theory would 
provide a similar insight into the nature of cosmological particle production. 
This, however, has turned out to be rather difficult, particularly because of 
problems of formulating string theory in time-dependent backgrounds.

Recent progress in two dimensional non-critical string theory
has improved the situation somewhat \cite{McGreevy:2003kb,Klebanov:2003km}. 
In this case there is a well understood holographic description --- the matrix quantum 
mechanics of open strings --- and also a closed string field 
theory, the  two dimensional collective field theory of the eigenvalue density
\cite{Das:1990ka}. 
The holographic description has no space and has a unique time, while, in 
the closed string description, space arises from the space of 
eigenvalues \cite{Das:1990ka, Polchinski:mf,Sengupta:1990bt}. 
Small fluctuations of the collective field then represent the 
perturbative states of the closed string theory (which in this case is a single massless 
scalar), while nontrivial time-dependent states, involving 
macroscopic numbers of decaying D0 branes, correspond to 
cosmological evolution. Such time dependent states can be studied 
unambiguously using the matrix model description.

The interpretation of time-dependent solutions in matrix theory as matrix cosmology
was introduced recently by Karczmarek and Strominger \cite{Karczmarek:2003pv} (the
solutions themselves have been known for some 
time \cite{Minic:rk,Moore:1992gb,Alexandrov:2002fh}). 
In the present paper we generalize their solutions by recognizing them as $W_\infty$ 
transforms of the ground state. This representation immediately allows the
generation of several infinite families of cosmological solutions. Although our 
derivation is classical the solutions clearly exist at the quantum level. 

The fluctuations around the cosmological backgrounds are efficiently described 
by the collective field theory of the matrix eigenvalues. The fluctuating field in fact
reduces to a massless scalar field in two dimensions, a popular toy model for 
studying vacuum ambiguities. The novelty in the present context is that the 
vacuum state of the scalar field is inherited from the ultraviolet completion of 
the theory, {\it i.e.} the matrix model. This means the short-distance divergences 
of the energy-momentum tensor are cancelled, rather than subtracted as in 
usual quantum field theory. It also means a specific, finite, energy is associated 
with the static vacuum, namely the energy which, in the matrix model, is the 
standard one loop energy of the ground state. In matrix cosmology this
ground state energy will contribute a novel time-dependent term in 
the EM-tensor which happens to precisely cancel the usual contribution from 
particle production. The VEV of the EM-tensor is thus {\it identical} in the initial
and final states of the cosmology, a highly unusual situation. 

It is often confusing what the correct observables are in time-dependent string 
theory. This concern seems particularly acute in matrix cosmology where the 
universe tends to have dramatic initial and/or final conditions, such as the 
complete disappearance of space-time at early or late times. In the context 
of quantum field theory the problem with such non-adiabatic evolution is 
that the particle concept is not useful. In matrix cosmology there is a 
preferred vacuum, {\it i.e.} a notion of no particle state which is 
universal, and applicable even as space-time is disappearing. This 
enable us to discuss cosmological particle in a setting that would be
difficult to analyze using ordinary quantum field theory. 

This paper is organized as follows. In section \ref{sec:matrix} we introduce the 
cosmological solutions to the matrix-model.  In section \ref{sec:boguliubov} 
we first discuss the causal structure and the observables of matrix cosmology.
We then construct the explicit modes of the fluctuating quantum fields and 
deduce the corresponding Bogulubov coefficients for particle production.
Finally, in section \ref{sec:anomalies}, we compute the energy-momentum
tensor of the model, using anomalies. 

\vspace{.5cm}
\noindent {\bf Note added:} as this paper was being prepared we received \cite{Karczmarek:2004ph} which has 
some overlap with the present 
work, particularly section \ref{sec:boguliubov}. However, the main points
of the works are different.

\section{Matrix Cosmology}
\label{sec:matrix}
In this section we introduce the matrix cosmologies of Karczmarek and 
Strominger \cite{Karczmarek:2003pv}, along with our generalizations. We discuss 
in turn the matrix model description, the collective field theory, and the fluctuations 
in collective field theory. 

\subsection{The Fermion Phase Space Picture}

The holographic description of two dimensional closed string theory is
in terms of singlet states of the quantum mechanics of a single
$N \times N$ hermitian matrix $M$ with some invariant potential
${\rm Tr}~V(M)$ which has a quadratic maximum. We will choose the 
value of $V(M)$ at this maximum to be zero. The dynamics can be entirely recast in 
terms of the eigenvalues $x_i (t)$ of the matrix $M$ and interpreted as
$N$ fermions with positions $x_i (t)$ in an external potential $V(x)$. The double 
scaling limit consists of tuning the coupling constants involved in the potential and 
taking $N \rightarrow \infty$ so that the Fermi energy $- \mu_F \rightarrow 0$
while the rescaled Fermi energy $\mu = \beta N \mu_F$ is held fixed.
In this limit the coordinates and the momenta of the fermions may be
rescaled such that the single particle Hamiltonian becomes the inverted
harmonic oscillator
\ben
h = - {1\over 2}{d^2 \over dx^2} - \half x^2~.
\label{eq:one}
\een
The Fermi energy of this rescaled problem is $-\mu$. 
Interpreting 
the double scaling limit as a continuum limit of string world-sheets one can identify 
the corresponding string coupling as $g_s=  {1\over \mu}$.
 
In the classical limit we can discuss the dynamics in terms of fermion 
trajectories in phase space $(x,p)$. In the ground state, the Fermi 
surface is
\ben
\half p^2 - \half x^2 = -\mu ~,
\label{eq:two}
\een
which implies that all states with $p^2 < x^2 - 2\mu$ are filled.
In this paper we will concentrate on the region $x < 0$. As is 
known now, the excitations of the Fermi sea in the $x > 0$ region 
simply represents a second massless scalar present in 
the 0B theory \cite{Takayanagi:2003sm}. 

The action corresponding to (\ref{eq:one}) has an infinite 
symmetry algebra, $W_\infty$ \cite{Avan:1991kq, Das:1991qb,
Witten:1991zd}, whose generators are given by
\ben
w_{rs} = \half e^{(r-s)t}(x-p)^r~(x+p)^s~.
\label{eq:three}
\een
The Hamiltonian is one of these charges, $h = - w_{11}$.
At the classical level the charges $w_{rs}$ satisfy the Poisson bracket algebra
\ben
\{ w_{rs}, w_{r's'} \}_{PB} = (rs'-sr')w_{r+r'-1,s+s'-1}~.
\label{eq:four}
\een
Our interest in $W_\infty$ as that it generates nontrivial solutions: starting with any 
classical solution, a new classical solution can be found by transforming with an
$W_\infty$ element. In particular a static Fermi sea like (\ref{eq:two}) can be 
transformed into a time dependent Fermi sea by using charges with $r \neq s$. 
We will concentrate on solutions generated in this way by charges $w_{r0}$ 
and $w_{0s}$. For these the {\em finite} transformations of $x$ and $p$
become
\bea
w_{0s} : ~~~x' & = & x + \lambda se^{-st}(x+p)^{s-1}~,~~~~~
p' = p - \lambda se^{-st}(x+p)^{s-1}~.
\label{eq:fiv}
 \\
w_{r0} : ~~~x' & = & x + \lambda re^{rt}(x-p)^{r-1}~,~~~~~~~
p' = p + \lambda re^{rt}(x-p)^{r-1}~,
\label{eq:five}
\eea
where $\lambda$ is the parameter of transformation.
Thus, starting from the static Fermi surface (\ref{eq:two}), we can
obtain an infinite set of exact solutions characterized by Fermi surfaces
\bea
w_{0s}:~~~ \half(x^2-p^2) + \lambda se^{-st}(x+p)^{s} & = & \mu ~. \label{eq:siks} \\
w_{r0}:~~~ \half(x^2-p^2) + \lambda re^{+rt}(x-p)^{r} & = & \mu ~.
\label{eq:six}
\eea

Of particular interest are the solutions obtained by the actions of
$w_{01}$ and $w_{10}$. In these cases the $W_\infty$ transformations (\ref{eq:fiv}-\ref{eq:five}) 
reduce to conventional {\em coordinate transformations} which are simply time 
dependent shifts of the coordinate. It is then clear that we can create a 
two-parameter family of solutions by combining the transformations 
$w_{01}$ and $w_{10}$ with arbitrary parameters
\begin{eqnarray}
x' & = & x + \lambda_- e^{-t} + \lambda_+e^{t}~,
\\
p' &=& p - \lambda_-e^{-t} + \lambda_+e^{t}~.
\end{eqnarray}
The resulting solutions
\ben
\half(x^2-p^2) + \lambda_- e^{-t}(x+p)+ \lambda_+e^{t}(x-p) =  \mu ~,
\label{KSsol}
\een
are in fact the solutions discussed by Karczmarek and Strominger \cite{Karczmarek:2003pv}
(up to a redefinition of the string coupling constant).

The more general cosmological solutions given by (\ref{eq:six}) also correspond 
to smooth Fermi surfaces. A nice way to map them is to introduce the coordinates 
$x_\pm =x \pm p$ in phase space. Then the solutions generated by $w_{0s}$ 
take the form
\ben
x_-x_+ + 2\lambda s e^{-st}x_+^s = 2\mu ~.
\een
We will consider the case of $\lambda > 0$.
After the further rescalings $y_- = x_-/2 \mu\alpha$ and $y_+ =\alpha x_+$ 
with the time-dependent factor
\ben
\alpha =  [{\lambda s \over \mu}]^{1/s}~e^{-t}  ~,
\een
the Fermi surface becomes
\ben
y_- = {1\over y_+}(1-y_+^s) ~.
\label{ftwo}
\een
Interestingly this surface is symmetric around $y=0$ in phase space only for 
even $s$, since only then $y_-(-y_+)=-y_-(y_+)$ . Concentrating as
usual on the region in which $y_+ <0$ we find, for odd $s$ a smooth Fermi surface interpolating 
between $y_-\sim1/y_+$ for small $y_+$ and $y_-\sim y_+^{s-1}$ for large $y_+$. 
Importantly, since the quantity ${dy_-\over dy_+}$ vanishes only once in this 
domain, there are no "folds" in the Fermi surface. This means it can be 
characterized by its intersection with a $x_+ ={\rm constant}$ line, {\it i.e.} by
the function $x_-=x_- (x_+)$. For even $s$, there is no zero of
${dy_-\over dy_+}$ and the Fermi surfaces actually cross over to
positive values of $y_-$. 
In this description the Euler equations for the 
Fermi surface are simply 
\ben
\partial_t x_\pm = \pm x_\pm \pm x_\mp {\partial x_\pm \over
\partial x_\mp} ~,
\een
which may be easily verified for (\ref{eq:siks}-\ref{eq:six}). Alternatively, we can parametrize 
the Fermi surface by the values of $p$ at the two intersections with the $x={\rm constant}$ 
line, $P_\pm (x,t)$. Then the Euler equations take the form
\ben
\partial_t P_\pm = x - P_\pm \partial_x P_\pm ~.
\label{Peuler}
\een
In the remainder of this paper we will primarily discuss the original solutions (\ref{KSsol})
of \cite{Karczmarek:2003pv}.

\subsection{Collective Field Theory}
So far our discussion has been in the open string language of matrix
quantum mechanics. From this point of view we have described some  
matrix configurations which are time dependent, but there has not been 
a notion of  "space'', and therefore no "cosmology" to discuss. The spatial 
coordinate $x$ is an emergent quantity which can be seen only after 
passing to the closed string description by introducing the density of 
eigenvalues, or the {\em collective field}  
\ben
\rho (x,t) = \sum_i \delta (x-x_i(t)) ~.
\label{cov}
\een
In the continuum limit it is convenient to trade the density $\rho(x,t)$ for
a scalar field $\phi(x,t)$ through $\rho (x,t) = \partial_x \phi (x,t)$. 
The action of the field $\phi(x,t)$ is
\cite{Jevicki:mb,jev1}
\ben
S =  \int dt dx\left[{(\partial_t \phi)^2 \over 2\partial_x \phi} 
- {\pi^2\over 6} (\partial_x \phi)^3 + ({1\over 2}x^2 - \mu)
\partial_x \phi\right] + 
\Delta S ~,
\label{eq:seven}
\een
where $\Delta S$ is the singular term 
\ben
\Delta S = {1\over 2}\int dt dx~\partial_x \phi~\left[\partial_x
\partial_{x'}\log |x-x'|\phantom{1\over 1}\hspace{-.3cm}\right]_{x=x'} ~,
\label{eq:eight}
\een
which is a part of the Jacobian of the change of variables (\ref{cov}) from eigenvalues $x_i(t)$ to
the density $\rho(x,t)$ \cite{Jevicki:mb}. The singular term $\Delta S$ contributes at higher
order in the loop expansion parameter $g_s=\mu^{-1}$, but it will nevertheless play 
a central role in what follows.

Any distribution of eigenvalues corresponds to a definite state of the
two dimensional collective field theory. However, for generic Fermi
surfaces such a state cannot be described as a classical solution of
the collective field theory because of the presence of folds or
disconnected pieces \cite{Polchinski:uq,Dhar:1992cs}. In fact, it can
represent states of the theory where quantum dispersions of fields are
of the same order as their classical expectation values
\cite{Das:1995gd, Das:2004rx}. However, for the time dependent Fermi
surfaces considered in this paper the profiles are quadratic --- there
are no folds --- and such Fermi surfaces can indeed be represented as
classical solutions of collective field theory.

The classical equation of motion following from the Lagrangean (\ref{eq:seven}) 
is
\ben
2\partial_t{\partial_t\phi\over\partial_x\phi} = \partial_x\left[ \pi^2 (\partial_x\phi)^2
+\left({\partial_t\phi\over\partial_x\phi}\right)^2 -
(x^2-2\mu)\right] ~.
\label{ceom}
\een
since we must ignore $\Delta S$ in the classical limit. The {\it ansatz}
\ben
\partial_x \phi_0 = {1\over\pi}P_0 (x,t)~,~~~~\partial_t \phi_0 = -
{1\over\pi}P_0 (x,t) F(t) ~,
\label{eq:nine}
\een
where 
\ben
P_0 (x,t) = {\sqrt{ (x- {\dot F})^2 - 2\mu}} ~,
\label{eq:eleven}
\een
solves the equation of motion (\ref{ceom}) for all functions $F(t)$. Imposing the 
consistency conditions $\partial_t\partial_x\phi_0=\partial_x\partial_t\phi_0$
on (\ref{eq:nine}) we find
\ben
{d^2 F (t) \over dt^2} = F(t) ~,
\label{eq:ten}
\een
and so
\ben
F(t) = \lambda_- e^{-t} - \lambda_+ e^{t} ~.
\label{Fdef}
\een
An alternative procedure that yields the solution (\ref{eq:nine}) with (\ref{eq:eleven})
and (\ref{Fdef}) is to verify that the profiles of the Fermi surface
\ben
P_\pm(x,t) = \pm P_0(x,t) +F(t) ~,
\een
satisfy the Euler equations (\ref{Peuler}). 

The two parameter family of solutions to (\ref{eq:ten}) given in (\ref{Fdef}) is 
identical to the matrix model solution (\ref{KSsol}) that was generated from 
the static solution $F = 0$ by the action of $w_{01}$ and $w_{10}$.
The explicit solutions for other $w_{0s},w_{r0}$ solutions can be also
obtained in principle, though they involve solutions of higher order 
algebraic equations.

The form of (\ref{eq:eleven}) restricts the $x<0$ branch of interest to
\ben
x < -{\sqrt{2\mu}} + {\dot F} ~.
\label{eq:elevena}
\een
The solution generated by $w_{01}$ has $F(t) = \lambda_- e^{-t}$
so, as $t = -\infty$, the Fermi surface is pushed to the region of large 
negative $x$. In perturbation theory, the 
collective field therefore does not support excitations 
for any finite $x$ --- there is no universe at $t = -\infty$. At finite times 
there is some allowed region of $x$ which is growing at $t$ increases
so that, as $t\rightarrow \infty$, the Fermi surface becomes the static 
solution (\ref{eq:two}), which is the usual universe of the two dimensional 
string around the ground state. The entire solution thus represents 
the {\em creation} of a universe. In a similar way, solutions generated 
by $w_{10}$ represent the {\em destruction} of a universe. We will 
henceforth concentrate on universe destruction.

It will be important for our considerations to introduce a lower bound 
$x_{\rm min}=-\Lambda$ in (\ref{eq:elevena}) which restricts the 
universe to the finite volume
\ben
-\Lambda < x < -{\sqrt{2\mu}} + {\dot F} ~.
\label{eq:elevenb}
\een
The origin of the infrared regulator is the potential $V(x)$ {\em
before} performing the double scaling limit. This leads to
$\Lambda\sim\sqrt{\beta N}\sim1/\sqrt{\mu_0}\to\infty$.  We should
therefore think of the volume of the universe as a quantity of order
$\sqrt{\beta N}$ in $x$ space.

The solutions of collective field theory which correspond to the more
general solutions generated by $w_{0s}$ or $w_{r0}$ are more
complicated to obtain, since these correspond to {\em nonlinear}
transformations of the collective field. Nevertheless, as mentioned 
above, they can be obtained by solving higher order equations in terms of
phase space variables and translating these solutions in the
collective field language.

\subsection{Backgrounds in Matrix model}

What is the meaning of these cosmologies in the original 
holographic theory, {\it i.e.} matrix quantum mechanics? This is 
an interesting question because in this case the holographic theory 
has no ``space'' at all and these cosmologies are encoded as specific 
time dependent configurations of the matrix $M(t)$. The Hamiltonian
for matrix quantum mechanics reads
\ben
H_M = {\rm Tr}[\Pi_M^2 -M^2] ~,
\label{vone}
\een
and the $W_\infty$ charges are given by
\ben
w_{rs} ={1\over 2}e^{(r-s)t}{\rm
  Tr}\left[(M-\Pi_M)^r~(M+\Pi_M)^s\right] ~.
\label{vtwo}
\een
If $|\mu\rangle$ denotes the ground state of this system, a cosmological
background generated by $w_{s0}$ is denoted by
\ben
|\lambda\rangle = {\rm exp}[i\lambda w_{s0}]|\mu\rangle ~,
\een
The expectation values of a typical invariant quantity may then be 
formally expressed as
\ben
\langle\lambda | {\cal O} |\lambda\rangle
 = \langle\mu | {\cal O}^\prime |\mu\rangle ~,
\label{vfour}
\een
where
\ben
{\cal O}^\prime = e^{-i\lambda w_{s0}}{\cal O}e^{i\lambda w_{s0}} ~.
\label{vthree}
\een
This computation is purely formal because the left hand
side of eq. (\ref{vfour}) is not meaningful because the state $|\lambda\rangle$ is not 
normalizable. States such as $|\lambda\rangle$ form a subset of the discrete states 
in the $c=1$ model \cite{Gross:1990js}. These, it is believed, should not be 
really be regarded as states in the spectrum of the original model. Instead,
they are interpreted as deformed backgrounds.The formal manipulation suggests
that the appropriate deformed Hamiltonian is
\ben
H' = H_M + \lambda se^{st}{\rm Tr}(M - \Pi_M)^s ~.
\label{vfive}
\een
In other words, the holographic interpretation of the cosmological background is
the {\rm modified} matrix model whose Hamiltonian is $H'$. This interpretation
is similar to the proposed description of two dimensional black holes in terms 
a matrix model deformed by a ${\rm Tr}M^{-2}$ potential \cite{Jevicki:1993zg}.

\subsection{Fluctuations}
\label{subsec:fluct}

The sole perturbative closed string excitation around an arbitrary classical
background is given by the fluctuation of the collective field around 
the corresponding classical solution of collective field theory $\phi_0(x,t)$
\ben
\phi (x,t) = \phi_0 (x,t) + {1\over{\sqrt{\pi}}} \eta (x,t) ~.
\label{eq:twelve}
\een
In our applications we identify the field $\eta$ with the spacetime tachyon
\footnote{In general leg-pole factors must be taken into account. These
result in a nonlocal redefinition of the field which will not be essential in our 
discussion.}. Inserting the expression (\ref{eq:twelve}) into the action (\ref{eq:seven}) we find the 
quadratic action of the fluctuations
\ben
S^{(2)} = \half \int dx dt~{1\over \pi \partial_x\phi_0}
\left[(\partial_t \eta)^2 -
\left( (\pi \partial_x \phi_0)^2 - ({\partial_t \phi_0 \over 
\partial_x \phi_0})^2\right)(\partial_x\eta)^2 - 2{\partial_t \phi_0 \over 
\partial_x \phi_0}\partial_t \eta
\partial_x \eta\right] ~.
\label{eq:thirteenx}
\een
This form of the quadratic action is valid for fluctuations around {\em any} classical solution 
$\phi_0(x,t)$. It is useful to interpret the action in terms of a massless scalar field propagating 
in a spacetime with metric
\ben
ds^2 = - dt^2 + {(dx +{\partial_t \phi_0 \over \partial_x \phi_0}dt)^2 
\over (\pi \partial_x \phi_0)^2} ~.
\label{eq:fourteenx}
\een
For the specific case of matrix cosmology (\ref{eq:nine}) we have
\ben
S^{(2)} = \half \int dx dt~{1\over P_0 (x,t)}
\left[(\partial_t \eta)^2 -
\left(P_0(x,t)^2 - F(t)^2\right)(\partial_x \eta)^2 + 2 F(t) \partial_t \eta
\partial_x \eta\right]~,
\label{eq:thirteen}
\een
and the background metric becomes
\ben
ds^2 = - dt^2 + {(dx - F(t) dt)^2 
\over P_0(x,t)^2} ~.
\label{eq:fourteenxa}
\een

The coordinate $x$ takes values in the interval (\ref{eq:elevena}). 
At the end-points of this space the field $\eta (x,t)$ satisfies {\em Dirichlet} 
boundary conditions. This follows from the fact that the integral of the full 
collective field $\int dx \partial_x \phi$ is the total number of fermions and 
therefore fixed\footnote{Strictly this argument implies only that 
$\eta(x_{\rm min})-\eta(-\sqrt{2\mu})$=0. That fermions are in fact prevented 
from leaking out in either end of the interval is clear prior to taking the 
double scaling limit, and this property is inherited by the scaled theory.}.

At the classical level the action (\ref{eq:thirteen}) is invariant under Weyl
rescalings and so it determines the metric only up to an overall conformal  
factor. In (\ref{eq:fourteenxa}) this factor was chosen so that spacetime is flat. 
This can be seen explicitly by transforming the spatial coordinate as
\ben
x = -{\sqrt{2\mu}}\cosh y + {\dot F}~,~~~~~~~\tau= t ~,
\label{cotrans}
\een
so that the metric becomes
\ben
ds^2 = -dt^2 + dy^2 ~,
\label{eq:minkow}
\een
and the boundary condition simply becomes a Dirichlet condition at $y = 0$.
In these coordiantes the problem thus returns to the static case.

In terms of the original coordinates $(x,t)$ we have a problem similar
to the moving mirror problem \cite{Birrell:ix,Carlitz:1986nh}, with
the mirror trajectory given by the upper limit of the coordinate range
(\ref{eq:elevena}).  In the moving mirror problem particle production
can be entirely rephrased in terms of the anomalous transformation of
the energy momentum tensor under the conformal transformation which
makes the mirror stationary.  The vacuum in static coordinates then
appears as a collection of particles in the frame where the mirror
moves. Particle production in matrix cosmology works similarly except that,
as we shall see in section 4.2, the vacuum is prescribed differently. 

In usual quantum field theory there is, in the absence of special
symmetries, no sense in which one vacuum is preferred over the other
--- the appropriate choice is determined by the nature of {\em
observers}. This vacuum ambiguity enters concrete computations through
the normal ordering prescription which is implemented at "equal time",
a notion that depends on the observer or, more precisely, on the
coordinate system. String theory is different because ultraviolet
divergences are absent and therefore physical observables, such as the
one loop free energy, are finite. The way this comes about in the
collective field theory description is that the singular term
(\ref{eq:eight}) acts as a counterterm that cancels all
divergences. The form of this singular term singles out a specific
coordinate system, by regulating physical quantities at equal matrix
time. This circumstance forces us to use the same time $t=\tau$ in the
original $(x,t)$ coordinates and the static coordinates $(y,\tau)$.
\section{Particle Production in Matrix Cosmology} 
\label{sec:boguliubov}
In this section we discuss the causal structure of the metric seen by the 
fluctuations and define notions of {\it in} and {\it out} modes accordingly. 
We derive the nontrivial Bogolubov transform relating the {\it in} and {\it out}  
modes, and we determine the corresponding spectrum of  particles. 
We consider for definiteness the {\it destruction} of a universe, {\it i.e.} a  
draining Fermi sea ($F=-\lambda e^{t}$).
\subsection{Causal Structure}
\label{subsec:causal}

A general problem in the study of time-dependent backgrounds is
the definition of proper physical observables. For example, it is often difficult 
to define an S-matrix, because there are no suitable {\it in} and {\it out}
regions. One might expect that matrix cosmology would suffer from this problem, 
since the entire spacetime disappears at late times. However, as we 
discuss now, this is fortunately not the case. 
\begin{figure}[!h]
\begin{center}
\leavevmode
\hbox{%
\epsfxsize=6.0in
\epsfysize=3.0in
\epsffile{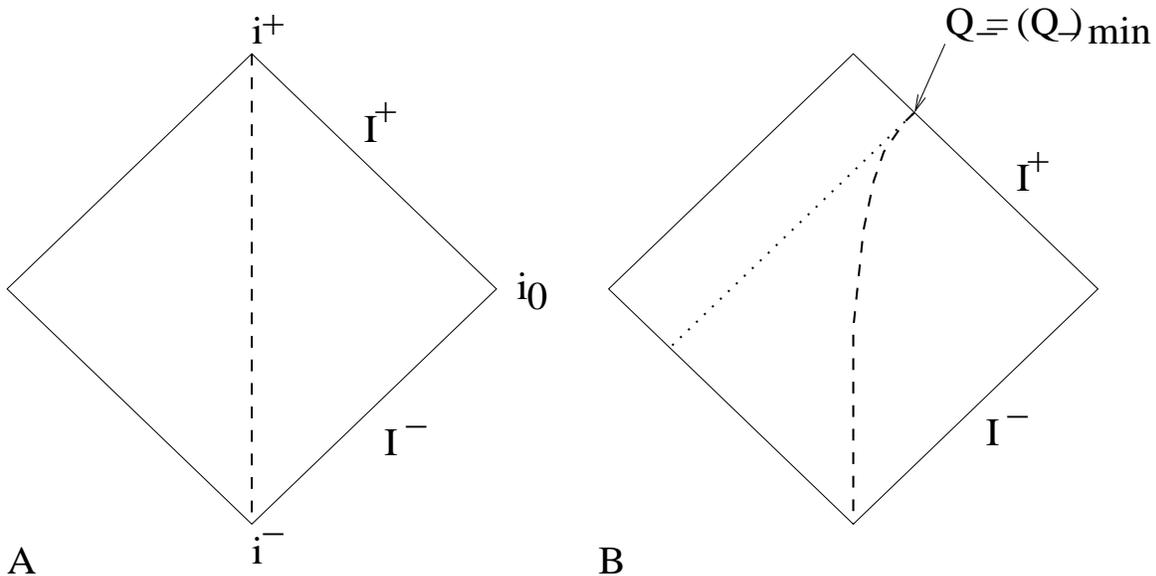}}
\caption{Causal structures in (A) static ($y$) and (B) cosmological ($Q$) coordinates. 
The dashed lines refer to the boundary of spacetime (the endpoint of the
eigenvalue distribution) in the two different coordinates.}
\label{causal}
\end{center}
\end{figure}

The most natural coordinate for spacetime processes is not the matrix model 
coordinate $x$, but rather the exponentiated coordinate $Q$ defined 
through\footnote{In the literature one often encounters  the coordinate 
$q$ defined through 
$x=-e^{-q}$ (see {\it e.g.} \cite{Polchinski:1994mb}). Our $Q\simeq q$ at asymptotic
distances but $Q$ has the advantage that the turning point of the Fermi sea is at $Q=0$.}
\ben
x =- \sqrt{2\mu} \cosh Q \label{Qdef}
\een
We will think of $Q$ as the physical spacetime coordinate and refer to it as
the cosmological coordinate. In our applications we will in addition find it essential 
to employ the static coordinate $y$ which transforms the matrix cosmology to 
Minkowski space (\ref{eq:minkow}) with a Dirichlet condition at $y=0$. 
According to (\ref{cotrans}) we have the relations
\ben
\cosh y = \cosh Q + {\dot{F} \over \sqrt{2\mu}} = \cosh Q - {\lambda \over \sqrt{2\mu}} e^t
\label{cotrans2}
\een
The time coordinate $t$ is the same for all these coordinate systems.

In the static coordinates ($y$) a natural {\it in} region ${\cal I}^-$ is defined by 
taking the light-cone coordinate $y_-=y-t\to\infty$. Similarly, a natural {\it out} 
region ${\cal I}^+$ is defined by taking the light-cone coordinate $y_+=y+t\to\infty$.
It is important to distinguish these regions from the future and past time-like infinities, ${\imath^{\pm}}$, which are reached by taking $t \to\pm \infty$ at fixed $y$,
and also from the space-like infinity, ${\imath^0}$, which is reached by taking 
$y\to + \infty$ at fixed time $t$. These are all indicated in figure 1(A).

This discussion of asymptotic regions in the flat coordinates translates nicely 
into the cosmological coordinates ($Q$). Indeed, define light-cone 
coordinates $Q_\pm=Q\pm t$ and use (\ref{cotrans2}) to obtain
\bea
Q_{\pm} & = & Q \pm t \nn \\
        & = & \pm t + \cosh^{-1} \left\{ \cosh y + {\lambda \over \sqrt{2\mu}} e^t \right\} \nn \\
 & = & \pm {{y_+ - y_- } \over 2} + \cosh^{-1} \left\{ \cosh \left({y_+ + y_-  \over 2} \right) + {\lambda \over \sqrt{2\mu}} e^{{y_+ - y_-} \over 2}  \right\} \label{Qy_pm}
\eea
The {\it in} region ${\cal I^-}$ is defined by taking $y_- \to \infty$ with $y_+$ fixed so 
(\ref{Qy_pm}) gives
\ben
Q_\pm = y_{\pm} + O(e^{-y_-}) ~~~~~~~~{\mathrm on}~{\cal I^-}
\een
The {\it in} region ${\cal I^-}$ is thus defined in physical coordinates by taking
$Q_- \to \infty$ with $Q_+$ fixed. Of course this is rather obvious, since it
is clear from (\ref{cotrans2}) that the static coordinates ($y$) agree with
the cosmological coordinates ($Q$) at early times. The point of making 
the notion of "early times" precise is to  avoid confusion
about the concept "late times" in the following. 

The {\it out} region ${\cal I^+}$ is defined by taking $y_+ \to \infty$ with $y_-$ fixed; 
and so (\ref{Qy_pm}) gives
\ben
Q_\pm = y_{\pm} + \log \left(1 + \lt e^{-y_-} \right) + O(e^{-y_+}) ~~~~~~~~{\mathrm on}~{\cal I^+} \label{Qminust}
\een
where $\lt= \sqrt{2\over \mu} \lambda$. Thus ${Q_+ \to \infty}$ with $Q_-$ fixed will 
take one to the {\it out} region ${\cal I^{+}}$. The relation (\ref{Qminust}) is easily 
inverted on ${\cal I^+}$ to find
\ben 
y_\pm = Q_\pm + \log \left( 1 - \lt e^{-Q_-} \right) +
O(e^{-Q_+}) \label{yminust} 
\een
An important feature of this expression is the branch cut at $Q_- = \log\lt$. Thus
we are reaching the {\it out} region by taking  
${Q_+ \to \infty}$ with $Q_-$ fixed {\it at some value larger than the minimal
value $(Q_-)_{\rm min} = \log\lt$}. The significance of this limiting value is that the 
boundary of the draining Fermi sea asymptotically approaches the light-like 
trajectory parametrized by $Q_- = (Q_-)_{\rm min}$. This is indicated on Figure 1(B).

In summary, the disappearance of spacetime by definition prevents 
the presence of a future timelike infinite ${\imath^+}$. However, there
is still a well-defined {\it out} region ${\cal I}^+$, defined by taking 
${Q_+ \to \infty}$ with $Q_-$ fixed at some value larger than the asymptotic
trajectory of the tachyon wall $(Q_-)_{\rm min} = \log\lt$. 
This means it makes sense to discuss observables in the form of S-matrix 
elements. 

\subsection{The {\it in} and {\it out} Modes} 
\label{subsec:inout}
The equation of motion for the fluctuating collective field $\eta$ can be solved 
exactly. This is most easily done in the static coordinates ($y$) where the
equation of motion is satisfied by simple plane waves. After taking the
Dirichlet boundary condition at $y=0$ into account, an obvious basis of
solutions is given by 
\ben
u^{\rm in}_\omega (y,t) = {1 \over \sqrt{\pi \omega}} e^{- i\omega t}\sin
\omega y~ \qquad \qquad (\omega > 0)
\label{confmode}
\een
and their complex conjugates. The normalization of (\ref{confmode}) has been 
chosen such that these modes form an orthonormal basis with respect to the 
Klein-Gordon norm
\ben
(u_{\omega'},u_\omega) = i\int_{\Sigma} d\Sigma^\mu \left( u_\omega \partial_\mu u^*_{\omega^{\prime}} - u^*_{\omega^{\prime}} \partial_\mu u_\omega \right) = 
\delta\left(\omega - \omega^\prime \right)
\label{kgproduct}
\een
on any Cauchy surface $\Sigma$. 
We can find the exact modes in the physical coordinates ($Q$) by solving (\ref{cotrans2}) 
for $y$ in terms of $Q$  and $t$, and substituting the result into (\ref{confmode}).
The modes obtained in this way will in general have a complicated dependence
on the time $t$, but this dependence become quite simple
in the {\it in} and {\it out} regions. 

In the {\it in} region $({\cal I^-})$ the static coordinates ($y$) coincide with the 
cosmological coordinates ($Q$) and so we can write (\ref{confmode}) as
\ben
u^{\rm in}_\omega(Q,t) = {1\over \sqrt{\pi\omega}} e^{- i \omega t} 
\sin \omega Q = {i\over \sqrt{4\pi\omega}} e^{i\omega Q_+} \label{inmodes-}~~~~~~~{\mathrm on}~{\cal I^-}
\label{confi+}
\een
In the second equality we omitted the term depending on $Q^-$ because, on 
${\cal I^-}$, this term does not contribute to the probability current. Equivalently,
the inner product (\ref{kgproduct}) reduces on ${\cal I^-}$ to
\ben
(u_{\omega'},u_\omega) = i \int dQ_+ \left( u_\omega \partial_{Q_+} u^*_{\omega^{\prime}} - u^*_{\omega^{\prime}} \partial_{Q_+} u_\omega \right)~~~~~~~~{\mathrm on}~{\cal I^-}
\label{innprod}
\een
According to (\ref{confi+}) the modes (\ref{confmode}) reduce to the standard 
positive frequency plane waves in the {\it in} region. It is for this reason that we 
have specified them from the outset by the superscript "in".  

In the {\it out} region $({\cal I^+})$ the static coordinate ($y$) is related to the 
cosmological coordinate ($Q$) through (\ref{yminust}) and so the modes (\ref{confmode})
take the form
\ben
u_\omega^{\rm in} (Q,t) =  -{i\over \sqrt{4\pi \omega}} e^{ i \omega Q_-} \left(1-\lt e^{-Q_-} \right)^{i \omega} ~~~~{\mathrm on} ~ {\cal I_+} \label{inmodes}
\een
We have omitted the term depending on $Q_+$ because this term does not contribute
to the current in the {\it out} region\footnote{This is clearest in the $y$-coordinates where
the inner product on ${\cal I}^+$ is written as in (\ref{innprod}) with $dQ_+\to dy_-$
and $\partial_{Q_+}\to\partial_{y_-}$. If we wish to write the inner product  
on ${\cal I}^+$ as an integral over $dQ_-$ we must use the more complicated
tangent derivative 
\ben
\left.{\partial\over\partial Q_-}\right|_{y_-}=\partial_{Q_-}
+\left.{\partial Q_+\over\partial Q_-}\right|_{y_+}\partial_{Q_+}
\een
where 
\ben
\left.{\partial Q_+\over\partial Q_-}\right|_{y_+}={\lt e^{-Q_-}\over1-\lt e^{Q_-}}~.
\een
}. The canonical modes (\ref{confi+}) in the {\it in} 
region thus evolve to the more complicated modes (\ref{inmodes}) in the {\it out} region. 
The change from dependence on $Q_+$ to dependence on $Q_-$ is due to
the reflecting boundary conditions on the field which turn left-movers into right movers. 

Since the {\it in} modes are rather complicated in the {\it out} region it is natural
to introduce a different basis which is simple there. The obvious choice is to
consider a set of modes parametrized by $\omega>0$ and which, 
in the {\it out} region, reduce to the canonical form
\ben
u^{\rm out}_\omega (Q,t) = -{i\over \sqrt{4\pi\omega}} e^{i\omega Q_-}\qquad \qquad~~~~{\mathrm on} ~ {\cal I_+}
\label{outmodes}
\een
It is obvious that modes in fact exist that satisfy the equations of motion 
everywhere and reduce to this expression in the {\it out} region: the {\it in}-modes and 
their complex conjugates all satisfy the equations of motion and, in the {\it out} region, 
they take the form given in (\ref{inmodes}) which, when the complex conjugates are 
included, span {\it all} functions of $Q_-$. The nontrivial content of selecting the 
modes (\ref{outmodes}) is the implied notion of positive frequency $\omega>0$
which, after quantization, amounts to the introduction of a particle concept. Since
spacetime is rapidly evolving at late times it is not obvious {\it a priori} that
any such notion should even exist in the {\it out} region. 
However, in the present context, the underlying matrix model singles out a preferred time 
coordinate $t$. The proposed {\it out} modes (\ref{outmodes}) are uniquely determined
by their canonical dependence on this time coordinate.  

\subsection{The Bogolubov Transformation}
\label{subsec:bogolubov}
The {\it in} modes depend on both the {\it out} modes and their complex
conjugates, as encoded in the Bogolubov transform,
\ben
u^{\rm in}_\omega (Q_-) =  \int_0^{\infty} d\bar{\omega} \, 
\left[\alpha(\bar{\omega}, \omega)
u^{\rm out}_{\bar{\omega}} (Q_-) + \beta(\bar{\omega}, \omega) 
\left(u^{\rm out}_{\bar{\omega}} (Q_-)\right)^* \right] ~.\label{bogo}
\een
We want to compute the Bogolubov coefficients $\alpha(\bar{\omega}, 
\omega)$ and
$\beta(\bar{\omega}, \omega)$. To do so we introduce the Fourier 
transform $F$
of the {\it in} mode through,
\ben
u^{\rm in}_\omega(Q_-) = \int_{-\infty}^\infty d\omega^\prime \,
e^{ i \omega^\prime Q_-} F(\omega ,\omega^\prime)~. \label{expand}
\een
We must be a little careful when inverting this expansion, because 
$Q_-$ has
a semi-infinite range bounded below by the minimum value $(Q_-)_{\rm 
min}=\ln\lt$. Accordingly we compute,
\bea
\int_{\ln\lt}^{\infty} {dQ_-\over 2\pi} \, e^{-i \bar{\omega} Q_-} 
u^{\rm in}_\omega(Q_-) &=&
\int_{\ln\lt}^{\infty} {dQ_-\over 2\pi} \, \int_{-\infty}^{\infty} 
d\omega^\prime \, e^{i (\omega^\prime - \bar{\omega})Q_-} F(\omega , 
\omega^\prime)\nn\\
&=& i \int_{-\infty}^{\infty} {d\omega^\prime\over 2\pi} \, {F(\omega 
,\omega^\prime) \over \omega^\prime - \bar{\omega}} e^{i (\omega^\prime 
- \bar{\omega}) \ln\lt}\nn\\
&=& F(\omega, \bar{\omega})~.
\eea
In the first step we assumed ${\rm Im}(\bar{\omega}) < 0$, which is 
implied already on the
left hand side to ensure convergence. In the next step we closed the
$\omega^\prime$-contour in the lower part of the complex plane.
Using (\ref{inmodes}) for the {\it in}-modes we find
\bea
F(\omega, \bar{\omega})  &=&  \int^{\infty}_{\ln\lt} {dQ_-\over 2\pi} 
\,  e^{-i \bar{\omega} Q_- }
u^{\rm in}_\omega (Q_-)\label{generalF}\nn\\
&=&-{i\over \sqrt{4\pi\omega}} \int^{\infty}_{\ln\lt}
{dQ_-\over 2\pi} \,  e^{i (\omega - \bar{\omega}) Q_-} \left(1 - \lt 
e^{-Q_-} \right)^{i\omega}
\nn\\
& = & -{i\over\sqrt{4\pi\omega}} {\lt^{i (\omega - \bar{\omega})}\over 
2\pi} \int_0^1 dz \, z^{-i(\omega-\bar{\omega})-1} \left( 1-z  
\right)^{i \omega} \label{FintQ} \nn \\
   & = & -{i\over\sqrt{4\pi\omega}} {\lt^{i (\omega - \bar{\omega})} 
\over 2\pi}B(i(\bar{\omega}-\omega), 1+i \omega )~.
\eea
We changed the integration variable to $z = \lt  e^{-Q_-}$. Comparing 
the definition
of the Bogolubov transform (\ref{bogo}) with the Fourier transform 
(\ref{expand}),
and recalling the definition of {\it out} modes (\ref{outmodes}), we now 
find
the Bogolubov coeffcients,
\bea
\alpha(\bar{\omega}, \omega) & = &  {1 \over 2 \pi} 
\sqrt{\bar{\omega}\over \omega}  \lt^{i {(\omega - 
\bar{\omega})}} B(i(\bar{\omega} -\omega),1 + i\omega)~, \\
\beta(\bar{\omega}, \omega) & = &  {1 \over 2 \pi} 
\sqrt{\bar{\omega}\over \omega}  \lt^{i {(\omega + 
\bar{\omega})}} B(- i(\bar{\omega} +\omega),1 + i\omega)~. \label{bog}
\eea

The non-vanishing of the $\beta(\bar{\omega}, \omega)$ is interpreted 
as particle
production in the matrix cosmology. The creation and 
annihilation operators
in the {\it in} and {\it out} vacua are defined by the expansions,
\bea
\eta(Q,t) &=& \int_0^{\infty} d\omega \left[\ai_\omega \ui_\omega(Q,t)  
+
\aid_\omega \uis_\omega(Q,t)  \right]~, \nn \\
&=& \int_0^{\infty} d\omega \left[\ao_\omega \uo_\omega(Q,t)  + \aod_\w 
\uos_\w(Q,t)  \right]~,
\eea
and so, comparing with (\ref{bogo}), one finds,
\bea
\ao_w = \int_0^{\infty} d \wb \left[\a(\w,\wb) \ai_{\wb} + \b^*(\w,\wb)
\aid_{\wb} \right]~.
\eea
The quantity that we want to compute is the expectation value of the
{\it out} number operator in the {\it in} vacuum,
\bea
N(w) &=& \langle \aod_\w \ao_\w \rangle_{in}\cr
&=& \int_0^{\infty} d \wb ~|\b(w,\wb)|^2\cr
&=& \int_0^{\infty} {d\wb \over 4 \pi (\w + \wb)}
{\sinh \pi \w \over \sinh \pi (\w+\wb) \sinh \pi \wb}~.
\eea
This is our final result for the particle spectrum on ${\cal I}^+$. The integral
is convergent at high energies, as it should be. There is a logarithmic divergence
for small ${\bar\omega}$ which is interpreted as a linear divergence in $Q_-$,
{\it i.e.} it is the energy {\it density} which is finite.

\section{The Energy Momentum Tensor}
In this section we analyze the energy-momentum tensor of the fluctuations
in the cosmological background. We begin by reviewing the static case and
then turn to the time-dependent setting.

\label{sec:anomalies}
\subsection{The Static Case}
We begin the discussion by considering the static case $F=0$. 
The Hamiltonian for the fluctuations then simplifies to
\ben
H =\int dx P_0 \left[ {1\over 2} \Pi_\eta^2 +  {1\over
    2}(\partial_x\eta)^2\right]+\Delta H ,
\label{hamfluc}
\een
where
\ben
\Delta H =\int dx P_0
\left[ {1\over 2\pi}\partial_x \partial_{x'}\log |x-x'|\phantom{1\over
    1}\hspace{-.3cm}\right]_{x=x'} ,
\label{hamjac}
\een
is the Hamiltonian form of  (\ref{eq:eight}). This last term is actually independent of the 
fluctuating field $\eta$, but it is included because it contributes at the same order in the loop 
expansion parameter $g_s=\mu^{-1}$ as the quadratic fluctuations. 
The measure $dx P_0$ that appear in both (\ref{hamfluc}) and (\ref{hamjac}) 
translates in the underlying matrix model to $de\rho(e)$, which is the obvious continuum
version of the sum over eigenvalues. The expressions in the square brackets then correspond 
to the quantum fluctuations of the eigenvalues. 

For explicit computations it is useful to employ the static coordinates $y$ where 
the metric for the fluctuation of the collective field is the simple Minkowskian 
(\ref{eq:minkow}), and the Dirichlet condition is imposed at $y=0$. In this coordinate
system the Green's function takes the simple form
\ben
D(\bar{t},\bar{y}; t, y) = -{1\over 4\pi}\log\left( {\Delta t^2 - \Delta y^2 \over \Delta t^2 - 
(y+\bar{y})^2 } \right), 
\label{greensfct}
\een
where $\Delta t =\bar{t}-t$ and $\Delta y = \bar{y} -y$. The denominator arises from 
the image charge that enforces the Dirichlet condition. The Green's function in other coordinate 
systems can be found by simply substituting expressions for $\tau$ and $y$ as functions of those 
other coordinates.

The Green's function diverges at coincident points so, to do the calculations, we must regulate
the theory. 
We interpret two point
functions by using a cutoff in $x$ coordinates according to the prescription 
\ben
\left\langle (\partial_x \eta)^2 \right\rangle = 
\lim_{\bar{x} \to x} \partial_{\bar{x}} \partial_x  D(\bar{t}, \bar{y} ; t, y) ,
\label{ptsplit}
\een
where the point-splitting $\bar{x} \to x+ {\epsilon \over 2};~~x\to x -{\epsilon \over 2}$ is
implied. The $t$ and $y$ (and their barred analogues) are functions of these slightly 
shifted $x$ (and ${\bar x}$). Evaluating the remaining expression we find  
\bea
\left\langle (\partial_x \eta)^2\right\rangle & = & 
-{1 \over 2\pi}  \lim_{\bar{x} \to x} \left\{ {\bar{y}^\prime y^\prime \over (\bar{y} - y)^2} + 
{\bar{y}^\prime y^\prime \over (\bar{y} + y)^2} \right\}\nn ,\\
 & = & -{1 \over 2\pi \epsilon^2} - {1\over 8 \pi} \left( {y^\p \over y} \right)^2 - 
{1\over 12\pi} \{ y, x \} + {\cal O}(\epsilon^2),
\label{etaxcorr}
\eea
where the limit $\epsilon \to 0$ is implied and we introduced the Schwarzian
derivative
\ben
  \{ y, x \} = {y^{\p\p\p} \over y^\p } - {3\over 2} \left( y^{\p\p}
    \over y^\p \right)^2 ,
\een
as well as the notations $y^\prime = \partial_x y=-1/P_0$ etc. (and similarly for ${\bar y}$ and ${\bar x}$). 

The canonical momentum is given by the Hamiltonian equation of motion as
\ben
\Pi_\eta = {1\over P_0}\partial_t\eta .
\label{pidef}
\een
Using this expression, it is straightforward to compute the two point correlator of the momenta 
using the Green's function (\ref{greensfct}) and the point-splitting procedure already used
in (\ref{ptsplit}). The result is 
\ben
\left\langle \Pi_\eta^2\right\rangle  =-{1 \over 2\pi \epsilon^2} + {1\over 8 \pi} \left( {y^\p \over y} \right)^2 - 
{1\over 12\pi} \{ y, x \} + {\cal O}(\epsilon^2) .
\label{momcorr}
\een
This result agrees precisely with (\ref{etaxcorr}), except for the sign of the second term, which is the part
that 
arises from the images. Indeed, the computations leading to the two results are almost identical because, 
after taking the $P_0$ appearing in the denominator of (\ref{pidef}) into account, temporal
derivatives act on the short distance part of the Green's function in the same way as spatial 
derivatives. This agreement between the potential energy (\ref{etaxcorr}) and the
kinetic energy (\ref{momcorr}) is referred to as the virial theorem. The origin of the virial
theorem in the matrix theory is the simple oscillator form of the potential after double scaling.

Collecting the results we find that the divergent pieces cancel
and the final result for the expectation value is 
\ben
H_{\rm gs} = -\int dx P_0 {1\over 12\pi}\{ y, x \} .
\een
of the Hamiltonian (\ref{hamfluc}) in the ground state. The explicit form (\ref{cotrans}) (with $F=0$) 
of $y(x)$ gives
\ben
\{ y, x \} = {1\over P^2_0}\left({1\over 2} +{3\mu\over x^2-2\mu}\right)\to {1\over 2P^2_0},
\label{yxsch}
\een
for large $x^2$. The extensive part of the energy is therefore
\ben
H_{\rm gs} =  - {1\over 24\pi} \left|\ln (-x_{\rm min})\right| =
{1\over 48\pi}{\rm ln}\mu_0 ,
\label{gsen}
\een
since $|x_{\rm min}|=\Lambda\sim 1/\sqrt{\mu_0}$. 
This result is interesting for several reasons:
\begin{enumerate}
\item
The singular term from the Jacobian (\ref{hamjac}) cancels the
regularized singularities from the two-point functions and so the
final result is finite, without the need for futher renormalization of
the collective field theory\footnote{This cancellation has been known from
early days of collective field theory
\cite{Das:1990ka, jev1, Andric:ck}. However, generally the image
term has been ignored in these computations and
our explicit verification of the virial theorem in the presence of regularization 
is also new, to the best of our knowledge.}. 
This explicit understanding of how the
matrix theory induces the correct counterterms in collective effective
field theory is the origin of a preferred vacuum in the theory.
\item
The image in the Green's function (\ref{greensfct}) contributes to the finite part of the
two-point functions (\ref{etaxcorr}) and (\ref{momcorr}), but these contributions 
cancel in the total energy. It is not surprising that boundary conditions are unimportant for the extensive part of the energy in the thermodynamic limit; but it is nice to see how it works explicitly.
\item
The finite part agrees with the one-loop result found by solving the matrix 
model explicitly 
\cite{Gross:1990ay}\footnote{The energy (\ref{gsen}) agrees with eq. 3.30 of \cite{Klebanov:1991qa} 
after dividing the result given there by two, because we only compute the 
energy of the $x<0$ part of the Fermi sea.}. This gives great confidence that 
we have interpreted the theory correctly.
\end{enumerate}
Let us also mention the further results:
\begin{enumerate}
\item
In collective field theory, the ground state of the matrix model is 
the vacuum defined in terms of the
mode expansion of the fluctuation field in terms of modes $e^{\pm i
  \omega t}\sin (\omega y)$ since this is the vacuum which leads to
the two point function used above.
\item
The presence of a finite ground state energy in this theory is an
important signature that we are dealing with a string theory. In fact, at
finite temperature $T$ this term, when added to the standard thermodynamic
contribution, leads to a T-dual answer symmetric under $\pi T
\rightarrow {1\over \pi T}$ characteristic of string theory \cite{Das:1990ka}.
\item
We have performed the calculation using a point splitting regulator in
$x$ space. If we use instead a regulator in $y$ space, the Schwarzian
derivative term will come from $\Delta H$. The divergences cancel as
before leading to the same finite answer.
\end{enumerate}

\subsection{The EM-tensor in Matrix Cosmology}
Let us now generalize these considerations to matrix cosmology. As 
we have emphasized, matrix cosmology is related to the static
case by a simple coordinate transformation which, near the 
asymptotic null infinity ${\cal I}^+$, takes the form
\bea
y_- &=& Q_- + \log[1-\tilde{\lambda} e^{-Q_-}] + {\cal O}(e^{-Q_+}),
\label{yminusta}
\eea
as $Q_+\to\infty$ with $Q_-$ fixed at some $Q_->\ln\lt$. 
As in previous sections we use the notation 
$y_\pm = y\pm t$ and $Q_\pm = Q\pm t$ for the static and 
cosmological light-cone coordinates and we also introduced the 
abbreviation 
$\tilde{\lambda} = \lambda {\sqrt{2 \over \mu}}$ to parametrize the 
limiting value $(Q_-)_{\rm min}=\ln\tilde{\lambda}$. 

The static coordinates ($y$) play a several roles in the problem. First, 
the cosmological coordinates ($Q$) reduce to the static coordinates
at early times; so the static coordinates define the {\it in} vacuum. 
Second, the Green's function (\ref{greensfct}) in static coordinates
transform simply into the cosmological coordinates, since the 
transformation (\ref{yminusta}) is conformal 
(it does not mix the $"+"$ and $"-"$ light-cone coordinates). This will allow 
us to use conformal techniques to study the problem at 
asymptotic null infinity ${\cal I}^+$. Following the standard strategy,
we begin the discussion by writing the results from the static case
in a form that transforms naturally under the conformal group. 

The extensive part of the energy was computed in the previous subsection 
with the result (\ref{gsen}). The volume factor in the static frame ($y$) is 
related to the infrared cutoff in the matrix model coordinates ($x$) 
through $x\sim e^y$ so we can express this result concisely
in terms of the energy density 
\ben
\epsilon=-{1\over 48\pi},
\een
in the static frame. Note that this formula is exact at ${\cal I}^+$ since the
subleading terms in (\ref{yxsch}) do in fact vanish as $Q_+\to\infty$ with 
$Q_-$ fixed. 
The classical expression for the pressure density is identical to that of the energy 
density so, repeating the arguments in the previous subsection, 
we find $p=\epsilon$. The complete energy-momentum
tensor in the static frame thus takes the form 
\ben 
\langle T_{y_+,y_+}\rangle = \langle T_{y_-,y_-}\rangle = - {1\over 48\pi},
\label{static}
\een
in light-cone coordinates 

Let us now transform to the cosmological coordinates $Q$ which contain 
the behavior at late times. Since (\ref{yminusta}) is a conformal 
transformation we can use the standard, anomalous, transformation 
rule 
\ben
T_{Q_-,Q_-} = ({\partial y_- \over \partial Q_-})^2 T_{y_-,y_-}+
{1\over 24\pi} \{ y_-, Q_- \}_S ,
\label{emanomali}
\een
of the EM-tensor. The second term is present because the 
short distance singularity of the EM-tensor, although still cancelled by the 
explicit counter-term (\ref{hamjac}) in the collective field theory, differs by 
finite amounts in the two coordinate systems. The expectation value of
(\ref{emanomali}) relates the EM-tensor of the static and the
cosmological coordinates. Evaluating the derivatives we find 
\bea
{\partial y_- \over \partial Q_-}&=& {1\over 1-\tilde{\lambda} e^{-Q_-}} ~,\\
\{ y_-, Q_- \}_S &=& {\lt e^{-Q_-}(1-{1\over 2}\lt e^{-Q_-})\over
(1- \lt e^{-Q_-})^2} ,
\eea
and then (\ref{emanomali}) becomes simply
\ben
\langle T_{Q_-,Q_-} \rangle= -{1\over 48\pi} .
\label{fourQ}
\een
This means physical observers will experience {\it no} outgoing flux of energy 
at ${\cal I}^+$. This result is a tremendous surprise. 

In standard computations, such as those considering moving mirrors, the static
EM-tensor $\langle T_{y_-,y_-}\rangle$ would be taken to vanish, and so
{\it only} the second term (the Schwarzian derivative) in (\ref{emanomali}) 
would contribute. The associated energy would be interpreted as the energy 
of the particles produced by the expansion. This is also the natural interpretation of the 
Schwarzian derivative here. However, in the present context
the particle production is partially obscured by another effect which is 
operative as well: the (one-loop) energy density of the "in" vacuum 
is non-zero and finite in the co-moving frame ($y$). The corresponding 
EM-tensor is blueshifted in the frame of asymptotic observers ($Q$) on 
${\cal I}^+$ and this leads to an energy flux. This effect is the origin of the first term in 
(\ref{emanomali}). Our remarkable result is that these two effects cancel 
each other precisely such that the cosmological vacuum has precisely 
the same EM-tensor as the static vacuum. This is very different from the standard 
computations. 

The EM-tensor (\ref{fourQ}) takes the same value along the entire ${\cal I}^+$
as it does on ${\cal I}^-$. This suggests that, despite appearances, ${\cal I}^+$
is a perfectly nice locus to define large classes of observables. One may
consider the scattering of particles from  ${\cal I}^-$ to ${\cal I}^+$, and
correlation functions on ${\cal I}^+$ itself. The existence of such observables
cannot be taken for granted in a time-dependent setting but, in matrix
cosmology, it would seem that they both exist and are computable.
It would clearly be interesting to make this more explicit.

Our result that the one-loop energy of the matrix cosmology agrees with that
of the static ground state is all the more surprising because the {\it classical}
energy in this specific background does receive the very specific time-dependent
contribution
\ben
E_{\rm cl}=E_{\rm cl}^{(0)} -{\dot F}\int_{-\Lambda}^{-\sqrt{2\mu}} dx x \sqrt{x^2-2\mu}
\een
where $E_{\rm cl}^{(0)}$ is the static ground state energy. This is consistent with 
(\ref{vfive}). That the one-loop energy is the same as in the ground
state of course means there is no energy flux.

In summary, we have found a remarkable cancellation between two apparently 
different contributions to the EM-tensor. The consequence is that the outgoing 
EM-tensor $T_{--}$ is identical in the {\it in} and {\it out} vacua. This is quite 
unusual and surprising, at least to us. It is natural to suspect that 
the cancellation ultimately stems from the integrable nature of the 
underlying matrix theory. To understand this better, it would be interesting to 
make the formal description of the matrix cosmology as a deformed matrix 
model more concrete. It would also be interesting 
to extend the computation of the EM-tensor to the broader class of cosmological 
solutions generated in section 2.2 from the $W_\infty$-symmetry of the model.

\section*{Acknowledgements}
We thank A. Jevicki,
J. Karczmarek,
H. Liu, L. Susskind and
T. Takayanagi
for discussions. PM would also like to thank A. Dhar, D. Ghoshal, R. Gopakumar,
D. Jatkar, G. Mandal, S. Minwalla, S. Mukhi, S. Naik, A. Sen and 
L. Sriramkumar for
useful discussions. A preliminary version of this work was presented
by P.M. in the National Workshop on String Theory, December 2003 held
at the Indian Institute of Technology, Kanpur, India. 
SRD and PM both thank the MCTP for hospitality during phases of this work. 
This work of J.D. and F.L. 
was supported in part by the Department of Energy under 
Grant No. DE-FG02-95ER40899.
The work of S.R.D. and P.M. was supported by 
National Science Foundation grant PHY-0244811 and the Department of
Energy grant No. DE-FG01-00ER45832.

\end{document}